\documentclass{baltzer}
\newcommand{\hba}{h}
\newcommand{\hq}{\hba{\!\!\!\!\mathchar'26\,}}

\input {epsfig.sty}

\begin{document}
\begin{frontmatter}  
%
\title{Muonium Spectroscopy}       
\author{Klaus P. Jungmann}      
\address{Physikalisches Institut, Universit\"at Heidelberg, D-69120 Heidelberg
\email{jungmann@physi.uni-heidelberg.de}}   
\runningauthor{Klaus P. Jungmann}
\runningtitle{Muonium Spectroscopy}
%
\begin{abstract}  
The electromagnetic interactions of electrons and muons 
can be described to very high accuracy within the framework of standard theory,
in particular within the hydrogen-like muonium atom. Therefore precision measurements
allow to test basic interactions in physics and to search for yet unknown forces.
Accurate values for fundamental constants can be obtained. 
Results from experiments on the ground state hyperfine structure 
and the 1s-2s intervals in muonium are described together with  
their relations to a new measurement of the muon magnetic anomaly.
\end{abstract}
%
\end{frontmatter}

\section{Introduction} 
To present knowledge leptons have dimensions of less than $10^{-18}m$ and may therefore 
be regarded as point-like objects. The muonium atom ($M=\mu^+e^-$) is the 
hydrogen-like bound state of leptons from two different particle generations,  
an antimuon($\mu^+$) and an electron($e^-$) \cite{Hug_90,Jun_99}. 
The dominant interaction within the M atom is electromagnetic and 
level energies can be calculated
in bound state Quantum Electrodynamics (QED) to sufficiently high 
accuracy for  modern high precision spectroscopic experiments. There are also 
contributions from weak interactions arising from $Z^0$-boson exchange and from
strong interactions due to vacuum polarization loops containing 
hadrons. They both can be obtained to the required level of precision using 
standard theory. In contrast to natural atoms and ions as well as
artificial atomic systems, which contain hadrons, M has the advantage that 
there are no complications arising from the finite size and the internal structure 
of any of its constituents. Precision experiments in M can therefore provide
sensitive tests of the standard theory and searches for new and yet unknown forces in nature.
Parameters of speculative theories, which try to expand the standard model in order to 
gain deeper insight into some of its not well understood features, can be restricted. 
In addition, fundamental constants like the muon mass $m_{\mu}$, its magnetic moment 
$\mu_{\mu}$ and anomaly $a_{\mu}$ and the fine structure constant $\alpha$ can be
obtained.   

\begin{figure}[t] 
\label{FIG1}
\begin{minipage}{2.5 in}
  \centering{
   \hspace*{-0.1in}
   \mbox{
    \epsfig{file=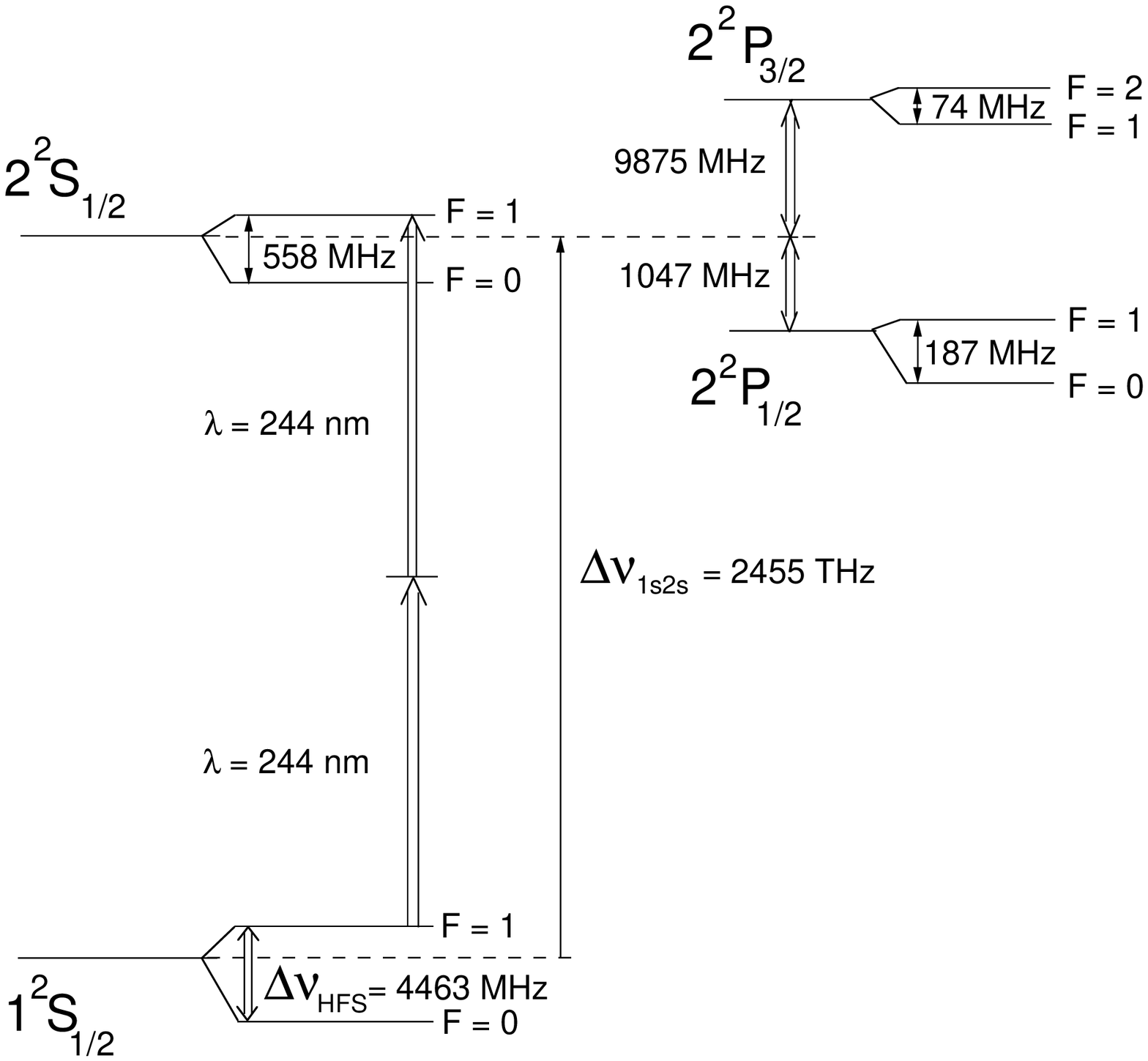,width=2.8in,clip=}
        } 
             }
\end{minipage}
 \hspace*{0.5in}
\begin{minipage}{2.5in}
  \centering{
   \hspace*{-0.15in}
   \mbox{
   \epsfig{file=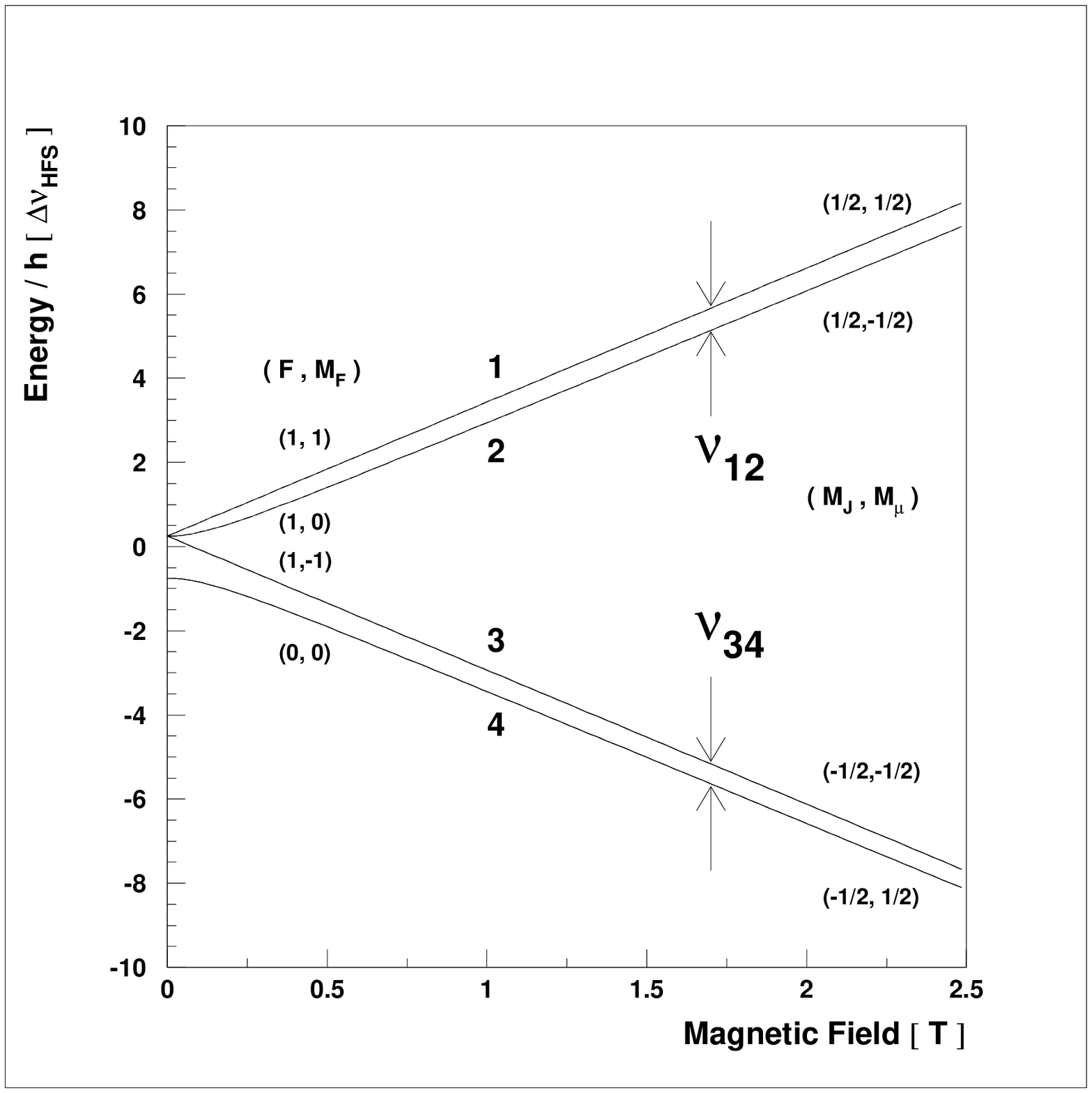,width=2.3in,clip=}
         }
             }
\end{minipage}
\caption[]{ Left: Muonium 
 n=1 and n=2 states.
 All indicated transitions could be induced to date.
 Right: Ground state Zeeman levels in an external magnetic field.
           }
\end{figure}

All high precision experiments in M up to date atom 
have involved the 1s ground state (see Fig.\ref{FIG1}),
in which the atoms can be produced in sufficient quantities \cite{Jun_99}. The most efficient 
mechanism is $e^-$ capture after stopping $\mu^+$ in a suitable noble gas,
where yields of 80(10)\% were achieved for Kr gas \cite{Hug_90}. This technique was used in
the most recent precision measurements of the atom's ground state hyperfine structure splitting
$\Delta \nu_{HFS}$ and $\mu_{\mu}$
at the Los Alamos Meson Physics Facility (LAMPF) in Los Alamos, USA
\cite{Liu_99}. 
Muonium at thermal velocities in vacuum can be obtained
by stopping $\mu^+$ close to the surface of a SiO$_2$ powder target, where 
the atoms are formed through $e^-$ capture and some of which diffuse through the target surface
into the surrounding vacuum. This process has an efficiency of a few \% and
was an essential prerequisite for Doppler-free two-photon laser spectroscopy
of the 1$^2$S$_{1/2}$-2$^2$S$_{1/2}$ interval $\Delta \nu_{1s2s}$ at the 
Rutherford Appleton Laboratory (RAL) in Chilton, United Kingdom \cite{Mey_99},
which yields an accurate value for $m_{\mu}$.
Electromagnetic transitions in excited states, particularly the 2$^2$S$_{1/2}$-2$^2$P$_{1/2}$
classical Lamb shift and 2$^2$S$_{1/2}$-2$^2$P$_{3/2}$ fine structure
splitting could be induced by microwave spectroscopy. However, because only moderate
numbers of atoms in the metastable 2s state can be produced with a beam foil technique, the
experimental accuracy is now the 1.5~\% level \cite{Ora_84,Bad_84},
which represents not yet a severe test of theory.

\section{Ground State Hyperfine Structure}

\begin{figure}[thb]
\label{FIG2}
 \unitlength 1.0cm
  \begin{picture}(15,7.2)  
   \centering{
   \hspace*{2.0cm}
   \mbox{
   \epsfig{file=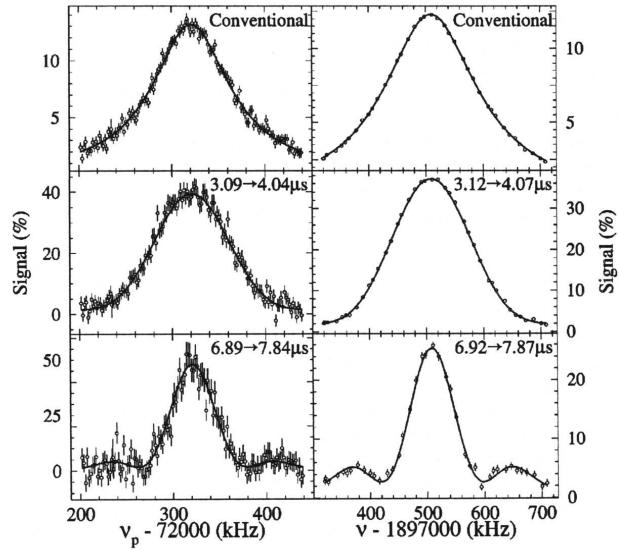,width=9cm,clip=}
         }
             }
  \end{picture}\par
  \caption[]{Samples of conventional and `old' M resonances at frequency $\nu_{12}$. 
  The narrow `old' lines are also higher. The lines in right column were recorded using
  a sweep of the magnetic field, which was measured in units of the proton NMR frequency
  $\nu_P$.
  The lines to the left were obtained using microwave frequency scans.}
\end{figure}

The most recent experiment at LAMPF
used a Kr gas target inside of a microwave cavity at typically
atmospheric density and in a homogeneous magnetic field of
1.7 Tesla. Microwave transitions
between the two energetically highest respectively two lowest Zeeman sublevels
of the n=1 state at the frequencies $\nu_{12}$ and $\nu_{34}$ 
(Fig.\ref{FIG1}) involve a muon spin flip. They
were detected through a change in the spatial distribution of $e^+$ from
$\mu^+$ decays, since due to parity violation in the $\mu^+$ decay
the $e^+$ are preferentially emitted in the  $mu^+$ spin direction.
As a consequence of the Breit-Rabi equation, which describes the
behaviour of the levels in a magnetic field, the sum of these frequencies equals
at any field value the splitting in zero field $\Delta \nu_{HFS}$
and their difference yields in a known field $\mu_{\mu}$.
The experiment utilized the technique of "old muonium", which allowed to reduce
the linewidth of the
signals below half of the "natural" linewidth $\delta \nu_{nat}=
(\pi \cdot \tau_{\mu})^{-1}$=145kHz, where $\tau_{\mu}$ is the muon lifetime
of 2.2 $\mu$ (Fig.\ref{FIG2}).
For this purpose an essentially continuous muon beam   
was chopped by an electrostatic kicking device into 4 $\mu$s long pulses with
14 $\mu$s separation.
Only atoms which were interacting coherently
with the microwave field for periods longer than several muon lifetimes
were detected \cite{Bos_95}.

The results are mainly statistics limited and improve
the knowledge of both $\Delta \nu_{HFS}$ and $\mu_{\mu}$
by a factor of three \cite{Liu_99} over previous measurements \cite{Mar_82}.
The zero field splitting is determined to
$\Delta \nu_{HFS}$=$ \nu_{12} + \nu_{34}$ = 4 463 302 765(53) Hz (12 ppb)
which agrees well with the theoretical prediction of 
$\Delta \nu_{theory}$= 4 463 302 563(520)(34)($\leq$100) Hz (120 ppb)~~\cite{Kin_98}.
Here the first quoted uncertainty is due to the accuracy to which the muon-electron mass 
ratio $m_{\mu}/m_e$ is known, the second error is from the knowledge of $\alpha$ as obtained
in electron g-2  measurements, and the third value corresponds 
to estimates of uncalculated higher order terms. The strong interaction contributes 250 Hz and 
a parity conserving weak interaction amounts to -65 Hz.
Among the possible exotic interactions which could contribute to 
$\Delta \nu_{HFS}$ is the conversion of muonium
to antimuonium, which is in the lepton sector an analogous  process to
the well known ${\rm K}_0$-$\overline{{\rm K}_0}$ oscillations in the quark sector. 
From a recent direct search at the Paul Scherrer Institute (PSI) in Villigen, Switzerland,
which itself could significantly restrict several speculative models, 
an upper limit of 9 Hz can be concluded for an expected line splitting
\cite{Wil_99,Jun_99}. Recently generic extensions of the standard model 
in which both Lorentz invariance and CPT invariance are not assumed have attracted widespread 
attention in physics \cite{Blu_98}.  Such models suggest diurnal variations of the ratio  
$({\Delta \nu_{12} - \Delta \nu_{34} })/({\Delta \nu_{12} + \Delta \nu_{34} })$ \cite{Kos_99} 
which are being searched for \cite{KJ_99}. 

The magnetic moment results from the measurements as 
$\mu_{\mu}/\mu_p$   = 3.183 345 24(37) (120 ppb) 
which translates into  
$m_{\mu}/m_e$  = 206.768 277(24) (120 ppb).
The hyperfine splitting is proportional to $\alpha^2 R_{\infty}$, with
the very precisely known Rydberg  constant $R_{\infty}$.
Comparing experiment and theory yields  
$\alpha^{-1}_{2}$= 137.035 996 3(80) (58 ppb) \cite{Liu_99}. 
If $R_{\infty}$ is decomposed into even more fundamental constants,
one finds  $\Delta \nu_{HFS}$ to be 
proportional to $\alpha^4 m_e/\hq$.
Using the value $\hq/m_e$ as determined in measurements of
the neutron de Broglie wavelength \cite{Kru_97} gives 
$ \alpha^{-1}_{4}$ = 137.036 004 7(48) (35 ppb).
In the near future a small improvement in $ \alpha^{-1}_{4}$ can be expected 
from ongoing determinations of $\hq/m_e$ in measurements of the 
photon recoil in Cs atom spectroscopy and a Cs atomic mass measurement.
The present limitation for accuracy of $ \alpha^{-1}_{4}$ 
arises mainly from the muon mass uncertainty. Therefore any better determination of the muon mass,
e.g. through a precise measurement of the reduced mass 
shift in $\Delta \nu_{1s2s}$, will result in an improvement of $\alpha^{-1}_4$.
At present the good agreement within two
standard deviations between the 
fine structure constant determined from M hyperfine structure
and the one from the electron magnetic anomaly is generally considered the
best test of internal consistency of QED, as one case involves bound state
QED and the other one QED of free particles.

\section{1s-2s Energy Interval}
Doppler-free excitation of the 1s-2s transition has been achieved in the past
at KEK in Tsukuba, Japan, \cite{Chu_88} and at RAL \cite{Maas_94}.
The accuracy of the latter measurement was limited by ac Stark effect
and a frequency chirp caused by rapid changes of the
index of refraction in the dye solutions of the amplifier stages
in the employed high power laser system.
A new measurement has been performed very recently at
the worlds brightest pulsed surface muon source at RAL
\cite{Mey_99}.
 
\begin{figure}[thb]
\label{FIG3} 
\begin{minipage}{2.5 in}
  \centering{
   \hspace*{-0.3in}
   \mbox{
   \epsfig{figure=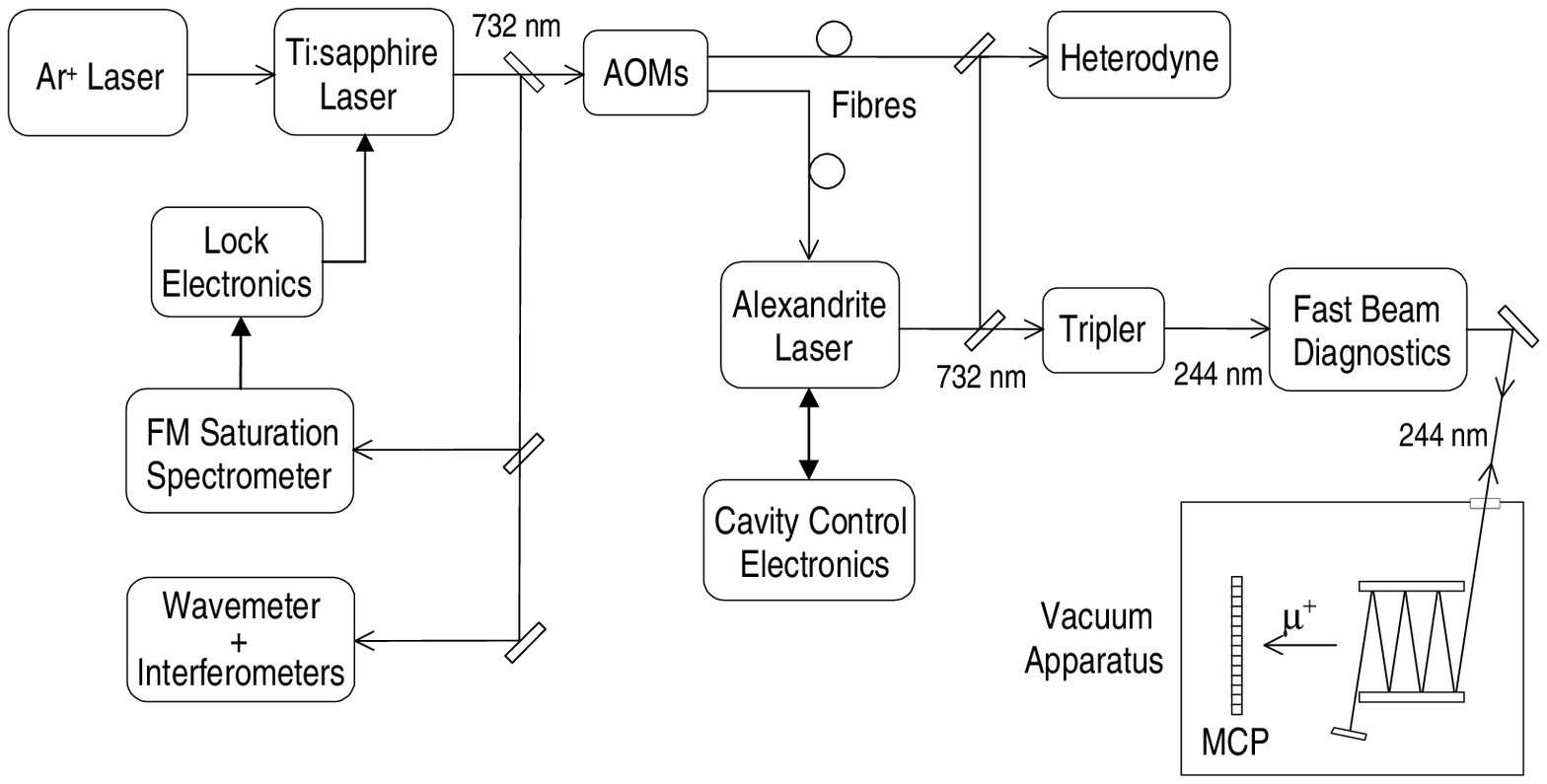,width=3.0in,clip=}
        } 
             }
\end{minipage}
 \hspace*{0.2in}
\begin{minipage}{2.5in}
  \centering{
   \hspace*{-0.0in}
   \mbox{
   \epsfig{figure=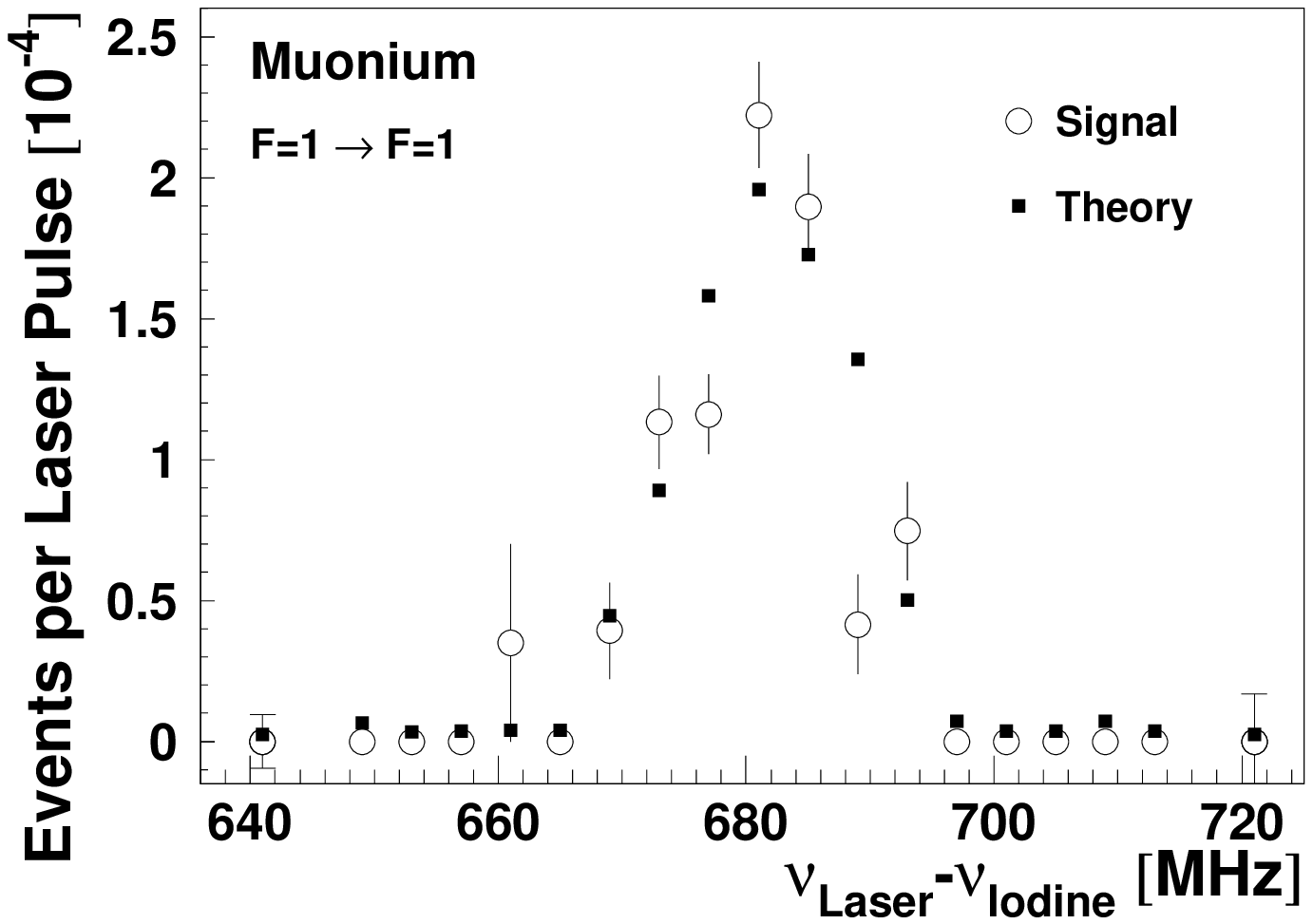,width=2.5in}
         }
             }
\end{minipage}
 \centering\caption[]
        {Left: Pulsed laser system in the M 1s-2s experiment.
         Right: Muonium 1s-2s signal.
         The frequency corresponds to the offset of the Ti:sapphire laser 
            from the iodine reference line. 
            The open circles are the observed signal, the
            solid squares represent the theoretical expectation 
            based on measured laser beam parameters
            and a line shape model {\protect \cite{Yak_99}}.
        }
\end{figure} 

The 1$^2$S$_{1/2}$(F=1) $\rightarrow$ 2$^2$S$_{1/2}$(F=1)
transition was induced when thermal muonium atoms interacted with the light field of 
two counter-propagating laser beams of wavelength 244~nm.
The two-photon excitation was detected by photoinization of the 2s state in the same light field.
The muons released thereby were identified and
counted. Their number as a function of laser frequency 
represents the experimental signal (Fig.\ref{FIG3}).  
The necessary high power UV laser light was generated
by frequency tripling the output of an alexandrite ring laser amplifier
in crystals of LBO and BBO. 
The alexandrite laser was seeded with light from a continuous wave
Ar ion laser pumped Ti:sapphire laser at 732~nm.
Fluctuations of the optical phase during the laser pulse (chirping) were compensated with
two electro-optic devices in the resonator of the ring amplifier to give
a swing of the laser lights frequency chirping  of less than about 5~MHz.
The fundamental optical frequency was calibrated by frequency modulation saturation
spectroscopy of the a$_{15}$ hyperfine component of the 5-13 R(26) line in thermally
excited $^{127}\rm{I}_2$ vapour which lies about 700~MHz lower than 1/6 of the
M transition frequency. It has been calibrated to 0.4~MHz 
\cite{Cor_99}.
The cw light was frequency up-shifted by passing through two
acousto-optic modulators (AOM's).

The experiment yields $\Delta \nu_{{\rm 1s2s}}$(expt.) = 2\,455\,528\,941.0(9.8)~MHz
in good agreement with a theoretical value
of $\Delta \nu_{{\rm 1s2s}}$(theory) = 2\,455\,528\,935.4(1.4)~MHz \cite{Pac_98}.
From these values the muon-electron mass ratio is found to be $m_{\mu^+}/m_{e^-}$ = 206.768\,38(17).
Alternatively, using $m_{\mu^+}/m_{e^-}$ extracted from the M hyperfine structure
experiment a comparison of $\Delta \nu_{{\rm 1s2s}}$(expt.)
and   $\Delta \nu_{{\rm 1s2s}}$(theory) yields the $\mu^+$-$e^-$
charge ratio as $Z= q_{\mu^+}/q_{e^-}=-1-1.1(2.1)\cdot 10^{-9}$. This 
is the best verification of charge equality in the first two generations of particles. 
The  existence of one single universal quantized unit of charge is solely 
an experimental fact for which no associated underlying symmetry has yet been revealed.
Gauge invariance assures charge quantization only within one generation of particles.

\section{Muon Magnetic Anomaly}

The muon magnetic anomaly $a_{\mu}$ is given, like in case of the electron,
mostly by photon and by electron-positron fields. However, the effects of heavier particles is
enhanced by the square of the mass ratio $m_{\mu}/m_e \approx 4 \cdot 10^4$.
The contributions of the strong interaction, which can be determined from a dispersion relation 
with the input from experimental data on $e^+$-$e^-$ annihilation into hadrons and hadronic 
$\tau$-decays, amounts to 58 ppm. The weak interaction adds 1.3 ppm. At present standard theory
yields $a_{\mu}$ to 0.66 ppm. Contributions from physics beyond the standard model
may be as large as a few ppm. Such could arise from, e.g., supersymmetry, compositeness
of fundamental fermions and bosons, CPT violation and many others.

A new determination of $a_{\mu}$ 
\cite{Car_99}
is presently carried out in a superferric  magnetic storage Ring 
\cite{Jun_98} at the Brookhaven National Laboratory
(BNL) in Upton, USA. It is a g-2 experiment in which the 
difference of the spin precession and the cyclotron frequencies
is measured.
In a first startup run, approximately the same 
level of accuracy for $\mu^+$ could be reached as the final result for this particle 
in a preceding experiment at CERN \cite{Bai_79}. Several technical improvements
were installed since, the most significant of which is a magnetic kicker, which allows to 
inject muons directly into the storage ring. This enhances the number of stored particles
by almost two orders of magnitude compared to the early stages of the experiment when 
the stored muons were born in the decays of injected pions.
Data have been taken which are expected to yield $a_{\mu}$ to 1~ppm. The data analyzed
so far have give the value with 5 ppm uncertainty. The value agrees 
with the prediction of standard theory.
The experiment aims for a final precision of 0.35 ppm. To be able to reach this goal, it
is essential to have $\mu_{\mu}$ to the 0.1 ppm level from muonium spectroscopy, since this
quantity is important in the extraction of the experimental result.

The experiment is planed for both $\mu^+$ and  $\mu^-$ as a test of CPT invariance.
This is of particular interest in view of the suggestion
by Bluhm et al. \cite{Blu_98} and Dehmelt et al. \cite{Deh_99}  
to compare tests of CPT invariance in different systems on a common basis, i.e.
the energies of the involved states. 
For measurements of magnetic anomalies this means that the
energies of particles with spin down in an external field need to be compared to
the energies of antiparticles with spin up.
The nature of g-2 experiments is such that they provide a figure of merit  
$r = |a^- - a^+| \cdot \frac{\hq\omega_c}{m \cdot c^2}$ for a CPT test, 
where $a^-$ and $a^+$ are the positive and negative particles magnetic anomalies, $\omega_c$
is the cyclotron frequency used in the measurement and $m$ is the particle mass.
For the past electron and positron measurements one has $r_e = 1.2 \cdot 10^{-21}$ \cite{Deh_99}
which
is a much tighter test than in the case of the neutral kaon system,
were the mass differences between $K^0$ and $\overline{K^0}$ yield 
$r_{K} = 1\cdot 10^{-18}$.
An even more stringent CPT test arises from the past muon magnetic anomaly  measurements
were  $r_{\mu} = 3.5 \cdot 10^{-24}$, which may therefore already be viewed as
the presently best 
%
%
\begin{table}[h]
\label{ACCEL}
 \caption[]{Muon fluxes of
   some existing and future facilities, Rutherford Appleton Lab\-orat\-ory (RAL),
   Japanese Hadron Facility (JHF), a new Neutron Spallation Source (NSS), 
   Muon collider (MC). }
 \begin{tabular}{|c|cccccc|}
 \hline
                    &RAL($\mu^+$)    &PSI($\mu^+$)   & PSI($\mu^-$)  &JHF($\mu^+$)$^\dag$
                    &NSS($\mu^+$)     &MC ($\mu^+$, $\mu^-$)\\
 \hline
  Intensity ($\mu$/s)& $3\times 10^6$ &$3\times 10^8$ &$1\times 10^8$ &$4.5\times 10^{11}$
                     &$4.5\times 10^7$ &$7.5\times 10^{13}$ \\
  Momentum bite   &&&&&&\\
  \hspace*{4mm} $\Delta$ pm/p[\%] & 10& 10            & 10            & 10
   & 10              & 5-10     \\
   Spot size     &&&&&&\\
   (cm $\times$ cm)         & 1.2$\times$2.0 &3.3$\times$2.0  &3.3$\times$2.0  &
   1.5$\times$2.0 &1.5$\times$2.0 & few$\times$few \\
   Pulse structure    & 82 ns & 50 MHz    & 50 MHz    & 300 ns & 300 ns & 50 ps\\
                    & 50 Hz & contin. & contin.&  50 Hz &  50 Hz & 15 Hz\\
\hline
 \end{tabular}
\end{table}
known CPT test based on system energies. With 
improvement expected in the BNL g-2 experiment one
can look forward to a 20 times more precise test 
of this fundamental symmetry.

\section{Future possibilities}

All precision M experiments are now limited by statistics. Therefore significant
improvements can be expected from either more efficient M formation, which
might in principle be possible to a small extent in the case of thermal M in vacuum.
The best solution, however, would be muon sources of higher intensities.
Such may become available  
in the intermediate future the Japanese Hadron Facility (JHF),
or the Oak Ridge (or a possible European) Spallation Neutron Source (NSS) 
Also the discussed Oak Ridge neutron spallation source.
The most promising facility is, however, a muon collider \cite{Palmer_98};
its front end will provide
muon rates 5-6 orders of magnitude higher than present beams (Table \ref{ACCEL}).

At such facilities there is in addition to 
more precise measurements in M
a variety of experiments on artificial atoms and ions like muonic
hydrogen and muonic helium which will allow to extract important parameters
describing the hadronic particles within these systems or fundamental
interactions, which could in no physical experiment thus far be accessed  
with suf\-fic\-ient precision for atomic, nuclear and particle theory \cite{Bos_96,Kaw_97}. 
It should be noted that new experimental approaches \cite{Wil_99,Jun_95} would 
also become feasible which might beneficially take advantage of, e.g., 
the time evolution of the atomic systems.

\section{Conclusions}
Although the nature of the muon - the reason for its existence - still remains a mystery,
both the theoretical
and experimental work in fundamental 
muon physics, have contributed to an improved understanding of basic particle interactions
and symmetries in physics.
Particularly muonium spectroscopy has verified the nature of the muon as
a point-like heavy lepton which differs only in its mass related parameters from 
the others. This fact is fundamentally assumed 
in every precision calculation within standard theory.
In addition, the measurements provide accurate values of fundamental constants.\\

\section{Acknowledgments}
The author wishes to acknowledge the work of the members of the 
different collaborations which produced the reported results.
This work was supported by The German BMBF, the German DAAD
and a NATO research grant.


%
\end{document}